\documentclass[12pt]{article}
\usepackage{epsfig}

\parskip=\medskipamount % space between paragraphs (TeX)
plus.3em minus.5em \abovedisplayshortskip=.5em plus.2em minus.4em
\renewcommand{\thesection}{\arabic{section}}

\catcode`@=11

\@addtoreset{equation}{section} \@addtoreset{equation}{subsection}
\def\theequation{\ifnum\value{section}=0 \arabic{equation}\ignorespaces
\else \ifnum\value{section}=-1 A.\arabic{equation}\ignorespaces
\else \ifnum\value{subsection}=0
\thesection.\arabic{equation}\ignorespaces \else
\thesection.\arabic{subsection}.\arabic{equation}\ignorespaces
                             \fi
                        \fi
                   \fi}

{\catcode`\'=\active \def'{{}^\bgroup\prim@s}}

\catcode`@=12

\newcommand{\bq}{\begin{equation}}
\newcommand{\be}{\begin{equation}}
\newcommand{\fq}{\end{equation}}
\newcommand{\ee}{\end{equation}}
\newcommand{\bqr}{\begin{eqnarray}}
\newcommand{\beqs}{\begin{eqnarray}}
\newcommand{\fqr}{\end{eqnarray}}
\newcommand{\eeqs}{\end{eqnarray}}

\def\bop#1{\setbox0=\hbox{$#1M$}\mkern1.5mu
    \vbox{\hrule height0pt depth.04\ht0
    \hbox{\vrule width.04\ht0 height.9\ht0 \kern.9\ht0
    \vrule width.04\ht0}\hrule height.04\ht0}\mkern1.5mu}
\def\Box{{\mathpalette\bop{}}}                        % box

\begin{document}
\thispagestyle{empty}

\begin{flushright}
\begin{tabular}{l}
% TEP- \\
\end{tabular}
\end{flushright}

\vskip .6in
\begin{center}

{\Large\bf  Program for IIB Derivative Corrections}

\vskip .6in

{\bf Gordon Chalmers}
\\[5mm]
% {\em address \\
%      address \\
% Los Angeles, CA } \\

{e-mail: gordon@quartz.shango.com}

\vskip .5in minus .2in

{\bf Abstract}

\end{center}

A Matlab program is presented that computes derivative corrections 
in the S-dual invariant formulation for IIB graviton scattering to 
any order in perturbation theory.  
The coefficients of the four-point function are produced, pertaining 
to the non-logarithmic terms.  The program can be modified to find 
coefficients of the higher-point functions.  Instantons have not 
explicitly been included.

\newpage
\setcounter{footnote}{0}

The graviton scattering amplitude in IIB superstring theory has 
been investigated recently in the context of manifesting S-duality.  
This manifestly S-dual graviton scattering has been analyzed systematically 
in \cite{Chalmers1}, and references therein.  

This text contains a computing program that automates 
the perturbative calculations.  To any order in derivatives this 
program computes the analytic portion of the graviton scattering, 
in the flat ten-dimensional superstring theory.  The D-instantons 
are allowed to be included, but their form is not placed in the 
program. 

The program is written in Matlab, and follows the procedure in 
\cite{Chalmers1} by partitioning numbers and attaching weighted 
trees to these numbers.  

Given the large number of papers on the topic, the author decided 
to place it on the archive.  The Matlab program may be 
found by downloading the TeX source.  

The quantum gravitational action may be numerically studied with 
this program, at the $\Box^k R^4$ level, with any set of integers 
$k$.

\vfill\break
\noindent{\it Matlab Program}

clear all

%%% These are the inputs for first part of program

Nmin=1; Nmax=10; 

trialnumber=200;

%%%

yp=0; flagno=1;

nullcase=0;

trialnumber=200

for i=Nmin:Nmax 

  for j=1:round(Nmax/3) 

    Oone(i)=0;  Partitiontype(i,j)=0; 

  end 

end

% loop to determine partitions 

for N=1:Nmax

for Np=1:round(Nmax/3);

 x(1,1)=0; p(1,1)=0; rp(1,1)=0; flagnumber(1)=0; 

 nullcase=0; a=0; dimp=0; flagno=0; timem=0; 
 
 totalflagno=0; v(1)=0; yp=0; 

 clear x,p,rp,flagnumber,nullcase; 

 clear a,dimp,flagno; 

 clear timem,totalflagno,v,yp;  

 clear i

if Np<=N

x(1,1)=0; clear x
p(1,1)=0; clear p

%% randomsampling to determine possible n_i in N=\sum n_i

%% the p matrix stores the values

 for j=1:trialnumber 

  for i=1:Np 

    x(i,j)=0; p(i,j)=0; 

  end 

 end

 for j=1:trialnumber

  flagno=1; timem=0; nullcase=0;

 while flagno==1

  yp=0;

 for i=1:Np 

  x(i,j)=2*round(((N-1)/2-1)*rand)+3; 

  yp=yp+x(i,j);  

 end 

  flagno=1; 

 if yp==N 

     p(:,j)=x(:,j);  

     flagno=0;

 end 

   timem=timem+1;  

   if timem>trialnumber

     flagno=0;  

     nullcase=1; 

   end

   if nullcase==1

     sp=0;  

     j=101;

   end 

 end
 
 end

% have to re-order the entries of p to avoid duplication such as 773 and 377  

if nullcase~=1

 for j=1:trialnumber

  for m=1:trialnumber

   for a=1:Np  

   for b=1:Np 

    if a<b 
 
     if p(a,j)>p(b,j)  

        test1=p(a,j); test2=p(b,j); 

        p(b,j)=test1; p(a,j)=test2;  

     end        

    end 

   end 

   end 

  end 

 end

%% the redundancy in p is eliminated 

%% values are stored in rp  

%% note that the zero vector is allowed 

v=size(p); dimp=v(2);

 for i=1:dimp 

  flagnumber(i)=0; 

 end

 for i=1:dimp

% redundancy check  

   for m=1:dimp

     if m==i 

        flagnumber(m)=1; 

     end

     if m~=i  

         if p(:,m)==p(:,i)  

           if i>m 

                flagnumber(m)=0;  

           else
                flagnumber(m)=1; 

           end 

         end 

      end 

   end 
  
 end

totalflagno=0;  

 for i=1:dimp 

   totalflagno=totalflagno+flagnumber(i); 

 end

rp(1,1)=0; clear rp

 for i=1:totalflagno

  for j=1:Np 

   rp(j,i)=0; 

  end 

 end

sp=0; k=0; 

 for i=1:dimp

   if flagnumber(i)==1  

      k=k+1; 

     rp(:,k)=p(:,i);  

     sp=sp+1;   
  
   end 

 end

%% the number of partitions is checked and stored in sp      
 
sp2=sp; test=0;

 for k=1:sp2 
 
      for l=1:Np  

        test=test+rp(l,k);  

      end 

 end 

sp=test/N

end 

end

%% sp is the number of partitions, this becomes the number of partitions

 for i=1:Np

  Partitiontype(1,i)=0; Partitiontype(2,i)=0; 

 end

    Partitiontype(N,Np)=sp;

end

   for j=1:Np

     Oone(N)=Oone(N)+Partitiontype(N,j); 

   end 

clear sp;

end

%% all partitions numbers are then stored in Partitiontype  

%%%%%%%%%%%%%%%%%%%%%%%%%%%%%%%%%%%%%%%%%%%%%%%%%%%%%%%%%%%%%%%%%%
%%%        	second part

% set of partitions stored in rp(i) via random sampling

% set of node weights stored in v(i,j) 

% set of choices found by expanding M in base 2  
%   coefficients in a(i), two choices per node

% perturbative contributions stored in Q(c) and A

% instanton=1 turns on the D-instantons, which have not been included 

%   a base three expansion is useful to separate the instantons 

%%%%%%%%%%%%%%%%%%%%%%%%%%%%%%%%%%%%%%%%%%%%%%%%%%%%%%%%%%%%%%%%%%
%%%%%%%%%%%%%%%%%%%%%%%%%%%%%%%%%%%%%%%%%%%%%%%%%%%%%%%%%%%%%%%%%%
%%%
%%%
%%%                       SECOND PART
%%%
%%%
%%%    Inputs are Nmin and Nmax, and the string coupling tau.   
%%%
%%%    The program uses Partitiontype to not compute when there 
%%%    is a 0 entry in the matrix.
%%%
%%%
%%%    The partitions of the integer N=\sum n_i are required, 
%%%    and the numbers n_i per partition were not saved in the 
%%%    first part of the program.
%%%
%%%    The derivative coefficients are stored in Amp(N,Np) and 
%%%    Coeff(N).
%%%
%%%    Trialnumber is an input for the random sampling.  
%%%

%%%   the coefficient c and A1 and A2 are required 
%%%    they have been set to 1  

%%%   these coefficients are in 0510233 equations (6) (16) (17) 

% a sum over N and and Np, stored in Amp(i,j)

%%%%%% these are the inputs 

Nmin=1; Nmax=10;

clear i

tau=i*1;  

tau1=real(tau);  tau2=imag(tau);

trialnumber=200;

A1=1;A2=1;

 for i=1:Nmax 

   tensorc(i)=1; 

 end

%% Partitiontype is also required

%%%%%%

for i=1:Nmax 

  for j=1:round(Nmax/3) 

     Amp(i,j)=0; 

  end 

end

for N=1:Nmax  

for Np=1:round(Nmax/3);

% this speeds up the algorithm 

   if Partitiontype(N,Np)~=0

%%%%%

  yp=0; flagno=1;

  nullcase=0;

% loop to determine partitions N and Np are inputs 

 x(1,1)=0; p(1,1)=0; rp(1,1)=0; flagnumber(1)=0; 

 nullcase=0; a=0; dimp=0; flagno=0; timem=0; 
 
 totalflagno=0; v(1)=0; yp=0; 

 clear x,p,rp,flagnumber,nullcase; 

 clear a,dimp,flagno; 

 clear timem,totalflagno,v,yp;

if Np<=N

 x(1,1)=0; clear x
 p(1,1)=0; clear p

%%%% the individual partition vectors are computed, stored in rp

 for j=1:trialnumber 

  for i=1:Np 

    x(i,j)=0; p(i,j)=0; 

  end 

 end

 for j=1:trialnumber

 flagno=1; timem=0; nullcase=0;

 while flagno==1

  yp=0;

   for i=1:Np 

      x(i,j)=2*round(((N-1)/2-1)*rand)+3; 

    yp=yp+x(i,j);  

   end 

     flagno=1; 

   if yp==N 

      p(:,j)=x(:,j);  

      flagno=0;

   end 

     timem=timem+1;  

   if timem>trialnumber

     flagno=0;  

     nullcase=1; 

   end

   if nullcase==1

     sp=0;  

     j=101;

   end 

 end
 
 end

% have to re-order the entries  to avoid duplication such as 773 and 377  

if nullcase~=1

 for j=1:trialnumber

  for m=1:trialnumber

   for a=1:Np  

   for b=1:Np 

    if a<b 
 
     if p(a,j)>p(b,j)  

        test1=p(a,j); test2=p(b,j); 

        p(b,j)=test1; p(a,j)=test2;  

     end        

    end 

   end 

   end 

  end 

 end

% eliminate redundancy

v=size(p); dimp=v(2);

 for i=1:dimp 

  flagnumber(i)=0; 

 end

 for i=1:dimp

% redundancy check  

   for m=1:dimp

     if m==i 

        flagnumber(m)=1; 

     end

     if m~=i  

         if p(:,m)==p(:,i)  

           if i>m 

                flagnumber(m)=0;  

           else
                flagnumber(m)=1; 

           end 

         end 

     end 

   end 
  
 end

  totalflagno=0;  

 for i=1:dimp 

   totalflagno=totalflagno+flagnumber(i); 

 end

  rp(1,1)=0; clear rp

 for i=1:totalflagno

  for j=1:Np 

    rp(j,i)=0; 

  end 

 end

  sp=0; k=0; 

 for i=1:dimp

   if flagnumber(i)==1  

      k=k+1; 

     rp(:,k)=p(:,i);  

     sp=sp+1;   
  
   end 

 end

%%% the above finishes computation of rp data set at given N,Np 

%%%% the quantum corrections are computed and stored in v(i,j)

%%%% the possible two choices at each node follows from a base 
%%%%  expansion in base 2 of all numbers up to 2^{nodes+1}

%%%% instanton 0 is a flag that turns off instanton corrections 

 instanton=0; 

%%%%

 w=size(rp)

%  base2N=2^(w(1)); 

 dimbase2N=w(1);

 Q=0; 

%%%% outputs are Amp(N,Np) per partition 
%%%% or Coeff(N) summed on partitions 

%%%% AmpNNp is the sum 

AmpNNp=0;

 for m=1:w(2) 

   R(m)=0;  T(m)=1;  

 end

 for i=1:dimbase2N 

   S(i)=0; zs(i)=0; 

 end

% sum over the rp vectors 

 for m=1:w(2)

  if rp(:,m)~=0

% store in the zeta values from the mth component of rp 

    for i=1:dimbase2N  

      zs(i)=rp(i,m); 

    end

% permutations for the weighted trees, 2 per node

  for A=1:base2N

    Q=A-1;

% base expand j; store in a(i), an array of size dimbase2N

   for l=1:dimbase2N  

    if l~=dimbase2N 

 %    P=mod(Q,2^dimbase2N*2^(-l));  

 %    a(dimbase2N-l+1)=(Q-P)/2^(-l)/2^dimbase2N; 

     Q=P;

    end 

    if l==dimbase2N 

     if Q==1 
     
       a(1)=1; 

     end  

     if Q==0 

       a(1)=0; 

     end  

    end 

   end

% k+3=N; k=(N-3)
% gmax=1/2(k+2)  k even, 1/2(k+1) k odd
% s=3/2+k/2

% conventions with \Box^k

k=N-3;  

s=3/2+(N-3)/2; 

 if mod(k,2)==0 

   gmax=1/2*(k+2); 

 end 

 if mod(k,2)==1 

   gmax=1/2*(k+1); 

 end

%%% perturbative corrections per Z_{s}, two of them

%%% instantons are not included yet, instanton==1 is the flag
 
  for i=1:dimbase2N 

   if a(i)==0     

%      S(i)=2*zeta(2*zs(i))/((3/2+k)*(3/2+k-1)-A1)*tau2^(2*zs(i)); 

          if instanton==1 

            S(i)=0; 
 
          end   

    end 

   if a(i)==1  

%      S(i)=2*pi^(-1/2)*zeta(2*zs(i)-1)*Gamma(zs(i)-1/2)/Gamma(zs(i))/((3/2+k-2*gmax)*(3/2+k-2*gmax-1)-A1)*tau2^(1-2*zs(i));

          if instanton==1 

            S(i)=0; 

          end 

   end 

     T(m)=T(m)*S(i);  

  end 

% end sum of the base expansion 

 end

% sum over the weighted graphs at a given mth rp vector 

 R(m)=R(m)+T(m);

% sum over all of the rp vectors   

% with conditionals 

end 

end 

end 

end

% sum over the partitions, or the rp vectors (and their weighted trees)

for c=1:w(2)  

  Amp(N,Np)=Amp(N,Np)+R(c); 

  AmpNNp=AmpNNp+R(c);  

end 

%%%%%%%%%%%%%%%

%%  close of Partitiontype conditional 
             
       end 

%%  close N and Np loop

   end 

end

% end sum on N and Np  

% sum per N 

for i=1:Nmax 

    Coeff(i)=0;

  for j=1:round(Nmax/3) 

    Coeff(i)=Coeff(i)+tensorc(i)*Amp(i,j); 

  end 

end 

%% the coefficients of \Box^k R^4 are stored in Coeff(N-3)  

\end{document}